\documentclass[prl,groupaddress]{revtex4}
\usepackage{epsfig}

\newcommand{\be}{\begin{equation}}
\newcommand{\ee}{\end{equation}}
\newcommand{\ba}{\begin{eqnarray}}
\newcommand{\ea}{\end{eqnarray}}

\begin{document}

\title{The slow dynamics of glassy materials: Insights from
computer simulations}

\author{Ludovic Berthier}
\affiliation{Laboratoire des Collo{\"\i}des, Verres
et Nanomat{\'e}riaux, UMR 5587, Universit{\'e} Montpellier II and CNRS,
34095 Montpellier, France}

\date{\today}

\begin{abstract}
The physics of glasses can be studied from many viewpoints, 
from material scientists interested in the development of new materials to 
statistical physicists inventing new theoretical 
tools to deal with disordered systems.
In these lectures I described a variety of physical 
phenomena observed in actual glassy materials, from disordered
magnetic systems to soft gels. Despite the very large gap between experimental
and numerical time windows, I showed that computer 
simulations represent an efficient theoretical tool which can shed
light on the microscopic origins of glassy dynamics. 
\end{abstract}

\maketitle

\hfill 
{\it Une garance qui fait entendre le violoncelle}

\hfill 
Vieira da Silva \\

Glassy states of matter continue to attract the interest 
of a large community of scientists~\cite{houches78,houches89,houches02}, 
ranging from material
physicists interested in the mechanical properties
of disordered solids, to theoretical physicists who 
want to describe at a more fundamental level the 
``glass state''~\cite{review,walterbook}. 
Glassy materials can be found in a variety
of materials, from soft matter (dense emulsions, concentrated colloidal
suspensions, powders) to hard condensed matter (molecular 
liquids, polymeric glasses, disordered magnets). 
Several glassy phenomena unrelated to specific materials, or even 
outside physics, are also discussed in this book.
A feature common to glassy materials is that their dynamics
gets so slow in some part of their phase diagram that they appear
as amorphous frozen structures on experimental timescales. 
The transition from 
a rapidly relaxing material (liquid, paramagnet...) to 
a frozen structure (window glass, spin glass, 
soft disordered solid...) is called  a ``glass transition''. 
For many glassy materials, a full understanding of 
the microscopic processes responsible for the formation of 
glasses is still lacking.

\begin{figure*}
\psfig{file=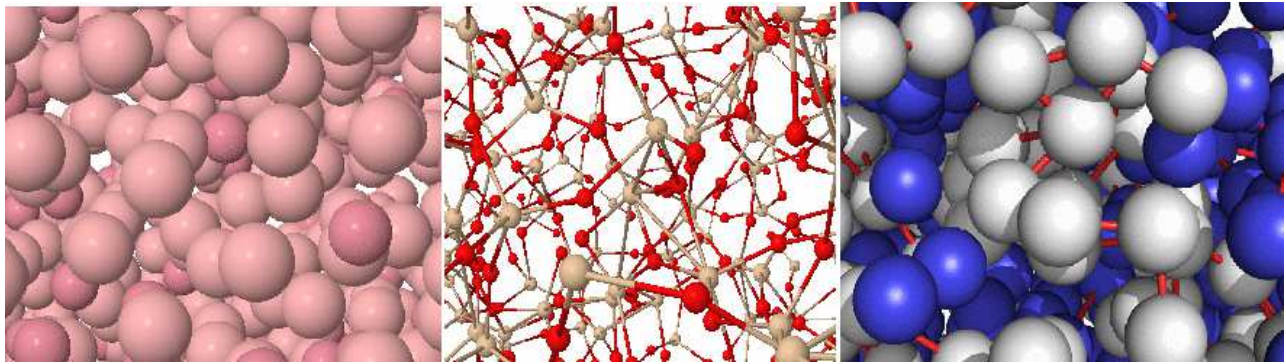,width=17cm}
\caption{Snaphots obtained from computer simulations of three different
materials characterized by glassy dynamics. Left: An equilibrium
configuration of a binary Lennard-Jones mixture, used as a model
system for metallic or colloidal glasses~\cite{KA}. Middle: Network 
structure of 
silica around 4000~K obtained from simulations of the
BKS model~\cite{beest90}.
Right: The picture shows oil droplets in white and blue, 
transiently connected by (red) telechelic polymers. For these
parameters, a system-spanning clusters of connected 
particles (highlighted in white) endows the system
with viscoelastic properties~\cite{pablo}.}
\end{figure*}

In Fig. 1 we present snapshots obtained from computer simulations
of three different models for materials characterized by slow
dynamics. The left panel shows a binary assembly of Lennard-Jones particles 
with interaction parameters specifically designed to avoid crystallization,
thus modelling either metallic or 
colloidal glasses~\cite{KA}. The middle panel is taken from numerical
simulations of a classical model for silica~\cite{beest90}, 
the main component of most
window glasses. The right panel shows the structure obtained 
in a soft gel~\cite{pablo} made
of oil droplets in water connected by telechelic polymers 
(long hydrophylic chains ended by small hydrophobic heads). 
In the three cases, the dynamics of individual particles can
get arrested on numerical timescales and the system essentially appears
as a disordered solid---a ``glass''. 
From a statistical physics point of view, two facts are 
quite puzzling. First, the structural properties of 
liquids and glasses are essentially indistinguishable. Second, 
there is no clear-cut phase transition between the two,  
so that the standard statmech language is not obviously the most
relevant one to describe the formation of these solids.

Just as in many different areas in physics, computer simulations 
are playing an increasing role in the field of glass 
formation~\cite{at,newman}. 
An obvious
reason is that, when simulating the materials shown in Fig.~1, the position
of all particles is exactly known at all times---the ultimate dream for an
experimentalist! Naturally, there are two immediate
drawbacks. Firstly, one might wonder
if it is really possible to simulate 
experimentally relevant materials in experimentally relevant 
conditions. Second question: What do we do with all this information? 

The answer to the first question is positive. 
With present day computers, it is possible to follow
for instance the dynamics of $N=10^3$ Lennard-Jones  particles
shown in Fig.~1 over 9 decades of time using about 3 months of
CPU time on a standard PC, thus covering 
a temperature window over which average relaxation 
timescales increase by more than 5 decades, quite a dramatic slowing down.
However, at the lowest temperatures studied, 
relaxation is still orders of magnitude faster than 
in experiments performed close to  the glass transition
temperature. Nevertheless, it is now possible to numerically access 
temperatures which are low enough that many features associated to the glass 
transition physics can be observed: strong decoupling 
phenomena~\cite{dec1}, clear  deviations from fits to
the mode-coupling theory~\cite{KA} 
(which are experimentally known  to hold only at high
temperatures), and crossovers towards activated dynamics~\cite{I}.
Of course, smaller timescales are accessed when simulating 
more complex systems, e.g. silica where Si and O atoms also carry 
charges and interact via a long-range Coulomb interaction, 
or more complex situations, e.g. boundary driven 
shear flows~\cite{varnik}, aging phenomena~\cite{jl}, 
or gel formation~\cite{dave}.   

The answer to the second question (what do we measure?) 
occupies the rest of this text. First one must make sure
that the glassy dynamics one seeks to study is at least
qualitatively reproduced by the chosen
numerical models, which are necessarily simplified
representations of the experimental complexity. One can for instance
devise ``theoretical models'', such as the Lennard-Jones liquid
shown in Fig.~1, which 
indeed captures the physics of glass-forming liquids~\cite{KA}.
One can also devise models inspired by real materials, such as 
the BKS model for silica and the connected
microemulsion shown in Fig.~1. 
The major signatures of glassy dynamics are indeed easily 
reproduced in simplified models and can therefore extensively be studied 
in computer simulations: slow structural relaxation, 
sudden growth of the viscosity upon lowering the temperature, 
aging phenomena after a sudden quench to the glass phase, 
non-Debye (stretched) form of the decay of correlation functions. 
Kob has given an extensive account of these phenomena in 
the proceedings of a previous school~\cite{walter}.

The important topic of dynamic heterogeneity, which emerged as an important
aspect of glassy materials during the 90s, 
is not covered in Kob's lectures, but alternative reviews exist~\cite{ediger}.
Although different phenomena usually go under the same
name, dynamic heterogeneity is generally associated to the existence, and 
increasing strength as dynamics gets slower, of non-trivial
spatio-temporal fluctuations of the local dynamical behaviour. 

Perhaps the simplest question in this context 
is as follows. On a given time window, $t$, particles in a liquid 
make the average displacement ${\bar d}(t)$, but  
the displacement of individual particles is distributed, $P(d,t)$. 
It is well established that $P(d,t)$ acquires 
non-Gaussian tails which carry more weight when 
dynamics is slower. This implies that relaxation in 
a viscous liquid must differ from that of a normal 
liquid where diffusion is Gaussian, and that non-trivial
particle displacements exist.
A long series of questions immediately follows this seemingly 
simple observation. Answering them
has been the main occupation of many workers 
in this field over the last decade.
What are the particles in the tails effectively doing? 
Why are they faster than 
the rest? Are they located randomly in space or do they cluster? 
What is the geometry, time and temperature evolution of the clusters?  
Are these spatial fluctuations correlated to geometric or thermodynamic
properties of the liquids? Do similar correlations occur in all
glassy materials? Can one predict these fluctuations 
theoretically? Can one understand glassy phenomenology
using fluctuation-based arguments? Can these fluctuations 
be detected experimentally? 

Although the field was initially principally driven 
by elegant experiments detecting 
indirect evidences of the existence of dynamic heterogeneity, 
and by a series of numerical observations in model liquids
or simplified glass models, theoretical progress
has been somewhat slower. It took some more time to realize
that dynamic heterogeneity could be studied using a set of well-defined 
correlation functions that can be studied either theoretically, in
computer experiments, or in real materials, thus allowing (in 
principle) a detailed comparison between theory and 
experiments~\cite{chi4,chi4theo}.

The main difficulty is that these correlators, unlike, say,
traditional scattering functions, usually involve more
than two points in space and time and represent therefore 
quite a challenge for computer simulations, but even more 
in experiments. To detect spatial correlations of the dynamics
one can for instance define ``four-point'' spatial
correlators, involving the position of two particles at two different
times, a quantity which can be directly 
accessed in simulations. Several such measurements have been 
performed, and directly establish that the dynamical slowing 
down encountered in glassy materials is accompanied by the 
existence of a growing correlation lengthscales over which
local dynamics is spatially correlated~\cite{chi4,chi4theo}. Together
with theoretical developments~\cite{rfot,Gilles,gc}, 
these results suggest that 
the physics of glasses is directly related to the growth
of dynamic fluctuations, similar to the ones encountered 
in traditional phase transitions~\footnote{Spin glasses
are one example where this behaviour is obviously realized
since three-dimensional spin glasses undergo
a genuine phase transition towards a spin glass phase
characterized by the divergence of a correlation length
measured via four-spin corrrelations, a static analog
of the four-point dynamic functions mentioned above.
During the school, however, it appeared that students did not 
seem to consider spin glasses as the most exciting example
of ``complex systems''.}.

Experimentally detecting similar multi-point
quantities in, say, a molecular liquid close the glass transition
would require having spatial resolution at the molecular level 
over timescales of the order of the second---a real challenge. 
Techniques have been devised to access these quantities 
in colloidal systems where microscopic 
timescales and lengthscales are more easily 
accessible~\cite{luca}.
Additionally, recent work has suggested that alternative 
multi-point correlation functions could be more easily studied 
in experiments, while containing similar physical 
informations~\cite{I,science}.

Despite being performed at lower temperatures 
and for liquids that are much more viscous than in
simulations, dynamic lengthscales measured in experiments 
are not much larger than in simulations, 
Typically, one finds that relaxation is correlated over 
a volume containing (at
most) a few hundreds of particles at low temperature.  
This means that even on experimental timescales, 
where dynamics is orders of magnitude larger than 
in numerical work, there is no trace of ``diverging''
lengthscales, as would be necessary for simple scaling   
theories to apply. Such modest lengthscales are, however, 
physically expected on general grounds: Because  
dynamics in glassy materials is typically 
thermally activated, a tiny change in activation energy
(possibly related to an even smaller growth of a correlation
lengthscale) translates into an enormous change in 
relaxation timescales~\cite{rfot,Gilles,gc,FH}. 

Although very few experimental results have been published, 
it seems that the dynamics of very many molecular liquids, and perhaps 
also of different types of glassy materials,  
could be analyzed along the lines of Ref.~\cite{science}, 
perhaps leading to a more complete 
description of the time and temperature dependences 
of spatial correlations in a variety of materials approaching the 
glass transition. It remains to be seen if these 
correlations can successfully and consistently
be explained theoretically, with precise predictions
that can be directly confronted to experimental results
with decisive results. 

I thank J.L.~Barrat, G.~Biroli, L.~Bocquet, J.P.~Bouchaud, D.~Chandler,
L.~Cipelletti, D.~El~Masri, J.P.~Garrahan, P.~Hurtado, R.~Jack, W.~Kob,
F.~Ladieu, S.~L\'eonard, D.~L'H\^ote, P.~Mayer, K. Miyazaki, D. Reichman, 
P.~Sollich, C.~Toninelli, F.~Varnik, S.~Whitelam, M.~Wyart, and 
P.~Young for pleasant and fruitful
collaborations on the topics described  during these lectures.
These notes were typed sitting next to a two week old little girl, 
whose non-glassy (but definitely complex) behaviour easily 
dominated over the noise of the laptop.

\end{document}